\documentclass[epj]{svjour}
\usepackage{latexsym}
\usepackage{graphicx}
\usepackage{dcolumn}
\usepackage{bm}
\usepackage{color}
\usepackage{amsmath}
\begin{document}
\title{Causality in thermoelectric systems: Insights from block diagrams}
\author{Y. Apertet \thanks{\emph{Email address:} yann.apertet@gmail.com} 
}%
%
%
\institute{Lyc\'ee Jacques Pr\'evert, F-27500 Pont-Audemer, France}
\date{Received: date / Revised version: date}
%
\abstract{
While Carnot's model engines demonstrate ideal performances regarding conversion efficiency, they cannot be actually used as energy converters since they are non causal systems. Such an unphysical behavior indeed restrains the working conditions to a single point where, in the case of a refrigerator (generator), the cooling power (output power) vanishes. Focusing on the example of a thermoelectric refrigerator, we study the impact of different dissipation sources on the causality of such systems. Basing our analysis on the block diagram description of this system, we discuss particularly the fact that heat conduction cannot ensure causality. 
} 

\maketitle

\section{Introduction}
In 1824, Carnot demonstrated in a seminal work \cite{Carnot1824} that heat transferred from a thermal reservoir at a temperature $T_h$ to a thermal reservoir at a lower temperature $T_c$ could be converted to work only with a limited efficiency. The maximum efficiency of such a conversion is the so-called Carnot efficiency $\eta_C$, given by
\begin{equation}
\eta_C = 1 - \frac{T_c}{T_h}.
\end{equation}
\noindent This optimal efficiency could however be reached only if the used engine operates reversibly. This reversibility condition is verified if the system does not demonstrate any dissipative process. Carnot thus proposed that the engine should work infinitely slowly. Consequently, although they are working with maximum efficiency, the so-called \emph{ideal Carnot engines}, i.e., without any dissipation, cannot provide any power, making them useless for practical applications \cite{Rebhan2002,Agrawal1997}. This ascertainment led to the emergence of the finite time thermodynamics in the middle of the twentieth century (see, e.g., Ref.~\cite{Andresen2011} for a recent review): Rather than focusing exclusively on efficiency maximization, it then appeared mandatory to also optimize power output even if it involves to consider dissipations as the system is no longer quasistatic. The optimization target was thus shifted from maximum efficiency to maximum output power. Special emphasis was placed on the efficiency at maximum power with the introduction of the paradigmatic Curzon-Ahlborn efficiency. This expression, derived by Curzon and Ahlborn in a very pedagogical article \cite{Curzon1975} (even if its paternity is still under debate as it was previously derived by several others researchers \cite{Vaudrey2014}), indeed is since then at the heart of numerous studies (see, e.g., Ref.~\cite{CalvoHernandez2014} and references therein).

While inclusion of dissipative processes in a heat engine description is often justified by practical considerations as we have just stressed, it might also be linked to theoretical considerations as discussed in Ref.~\cite{Ouerdane2015}, in particular to causality issue. In this article, we define a causal system as a system where the output at any time depends only on values of the input at the present time and in the past \cite{Oppenheim1996}. From this perspective, an ideal Carnot engine should be considered as a non-causal system since time's arrow does not apply to it due to the condition of vanishing entropy production \cite{Eddington1928}: It is not possible to associate a present time nor a past with this perfect engine. 

Beyond the condition of non-anticipation, the previous definition of a causal system also stresses the importance of the dependence between the input and the output of the system. The system will be considered as physical only if a constant finite driving force gives rise to a constant finite response \cite{DeGroot1984}. Causality thus may be viewed as the ability to reach a particular working point thanks to the load.
Just as an ideal electrical capacitance needs a dissipative electrical resistor to ensure the potential continuity when connecting to a perfect voltage source, a Carnot engine needs dissipations in order to ensure potential continuity at its edges, to connect with its load and, hence, to be driven by this latter. Without dissipations, ideal Carnot engine actually cannot be driven by an external input: It might be viewed as an isolated system that cannot exchange energy or matter with its environment. Thereby, an ideal Carnot engine is non causal since it possesses only a single working point where power vanishes. One may also state that Carnot engine straddles the fence, stranded between generator and refrigerator regimes.

\begin{figure}
	\centering
		\includegraphics[width=0.45\textwidth]{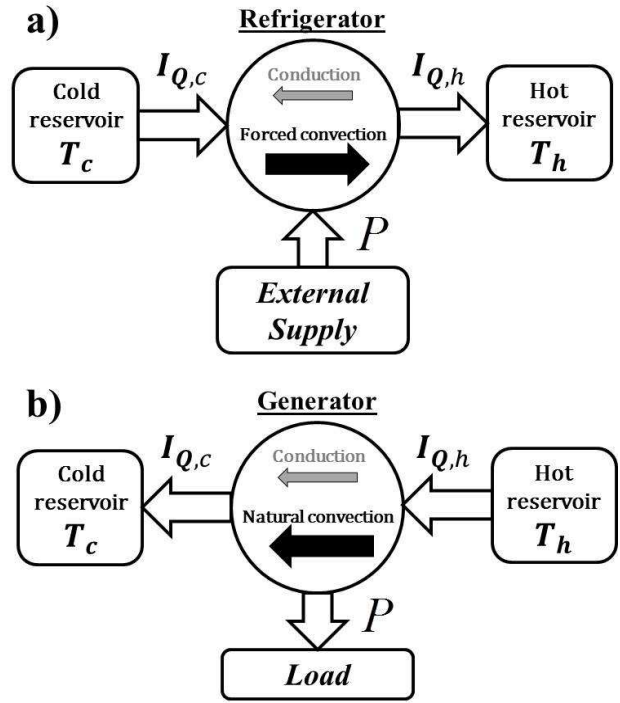}
	\caption{Schematic description of an autonomous heat engine in (a) refrigerator regime and (b) in generator regime. }
	\label{fig:figure1}
\end{figure}

All dissipation sources are however not equivalent when it comes to ensuring causality, i.e., to make the system depend on an external load. For example, while the introduction of dissipative thermal contacts between the system and the thermal reservoirs solve the causality issue as demonstrated by Curzon and Ahlborn \cite{Curzon1975}, this is not the case for thermal conduction inside the system. It was even demonstrated that these parasitic heat leaks should be avoided in order to get the maximum efficiency at maximum power. This assumption of vanishing thermal conduction is known as the \emph{strong coupling assumption} \cite{Kedem1965,VandenBroeck2005}. This article thus aims at clarifying the status of the main dissipations sources focusing on the particular relation between the system and its load. To that extent, rather than considering a cyclic engine as Carnot did, we focus on autonomous engines, i.e., engines working with steady-state conditions imposed by an external load. Such systems are schematically described on Fig.~\ref{fig:figure1}. It should be noted that for these systems, the transport of heat associated with energy conversion might be related to the movement of a \emph{working fluid} (in a broad sense as it might designate a genuine fluid but also an optical cavity mode for example \cite{Gelbwaser2015}). This useful heat transport could thus be described as a \emph{convective} process. When the system is used as a refrigerator, the external load supplies work in order to impose a heat flux from the cold reservoir to the hot reservoir (Fig.~\ref{fig:figure1}.a). Note that for a refrigerator the energy conversion effectiveness is evaluated through the coefficient of performance $\varphi$ defined as the ratio between the heat flux extracted from the cold reservoir and the power used to do it.
Conversely, in the generator regime, the natural convection of the working fluid leads to deliver work to the load. In both cases, an additional parasitic heat flux independent of the working fluid displacements could also appear: It corresponds to heat conduction inside the system.

To illustrate our point, we consider more specifically a thermoelectric system. This autonomous heat engine indeed appears to be a touchstone for irreversible thermodynamics as stressed by de Groot \cite{deGroot}. It allows to clearly identify each independent dissipation sources and characterize them through a single coefficient: The dissipation resulting from the displacement of the working fluid, corresponding to Joule heating in the thermoelectric case, is associated with the electrical conductivity $\sigma$ while the dissipation resulting from heat conduction is associated with the thermal conductivity $\kappa$. Moreover, the thermoelectric heat engine has been chosen as an example because it also displays some universality. It has indeed been demonstrated that other autonomous heat engines such as the Feynman ratchet have a behavior similar to thermoelectric engines and that they could be described using similar equations \cite{Apertet2014}. It is thus easy to extrapolate insights gained from the study of thermoelectric systems to other systems.

The thermoelectric figure of merit $ZT$ encompasses in a single variable the degree of dissipation inside the system \cite{Ioffe1957}. This figure of merit is defined by
\begin{equation}\label{ZT}
ZT = \frac{\sigma \alpha^2 \overline{T}}{\kappa},
\end{equation}
\noindent where $\overline{T} = (T_h + T_c)/2$ is the average working temperature of the system and $\alpha$ is the Seebeck coefficient reflecting the coupling between the electrical current and heat flux. The maximum attainable efficiency of a thermoelectric generator is directly linked to this figure of merit through the following relation \cite{Heikes1961}:
\begin{equation}\label{etamax}
\eta_{\rm max} = \eta_C  \frac{\sqrt{1+ZT} - 1 }{\sqrt{1+ZT} + T_c / T_h}.
\end{equation}
\noindent The maximum coefficient of performance of a thermoelectric refrigerator is also related to this parameter as
\begin{equation}\label{phimax}
\varphi_{\rm max} = \varphi_C  \frac{\sqrt{1+ZT} - T_h / T_c }{\sqrt{1+ZT} + 1},
\end{equation}
\noindent where $\varphi_C = T_c / (T_h - T_c)$ is the ideal coefficient of performance equivalent to the Carnot efficiency for the refrigerators. This ideal value corresponds to the behavior of a Carnot engine working in the refrigerator regime. As discussed by Littman and Davidson \cite{Littman1961}, $ZT$ can in theory reach infinite value and so, from Eq.~(\ref{etamax}), thermoelectric systems could in principle reach Carnot efficiency. However, an infinite figure of merit corresponds to two distinct situations regarding dissipations: Either the thermal conductivity $\kappa$ vanishes, i.e., there is no heat leaks through conduction, or the electrical conductivity $\sigma$ is infinite, i.e., there is no dissipation due to Joule heating even when there is an electrical current. Obviously, we can also add the most ideal situation where these two situations are met simultaneously. The former situation, $\kappa = 0$, is however believed to be a more suitable way to obtain a reversible thermoelectric engine \cite{Humphrey2005} also known as the \emph{best thermoelectric} \cite{Mahan1996}. We will stress in this article that the case $\sigma \rightarrow \infty$ is not a viable solution regarding causality.

The article is organized as follows. In Sec.~\ref{TEdescription}, we describe the thermoelectric refrigerator from a thermodynamic viewpoint and we give the corresponding modeling in the block diagram framework. Then, in Sec.~\ref{dissipation}, we discuss the benefits of the different dissipation sources, i.e., the parasitic heat conduction, the Joule heating and the finiteness of thermal contact conductances. We end this article with some concluding remarks.

\section{\label{TEdescription} Thermodynamical description of the thermoelectric conversion}

\subsection{Global description}

We consider an uniaxial homogeneous system in contact with a cold reservoir at temperature $T_{\rm c}$ and with a hot reservoir at temperature $T_{\rm h}$ and we choose to adopt, without loss of generality, the notations associated with the refrigerator regime as depicted in Fig.~(\ref{fig:figure1}.a) (working conditions of the others regimes are discussed in the Appendix). The heat currents exchanged with each of these reservoirs are then respectively given by \cite{Goldsmid1964}: 
\begin{equation}
I_{Q,c} = \alpha T_{\rm c} I - K_0 \Delta T - \frac{1}{2}R I^2,
\label{iqc}\end{equation}
\noindent and
\begin{equation}
I_{Q,h} = \alpha T_{\rm h} I - K_0 \Delta T + \frac{1}{2}R I^2,
\label{iqh}\end{equation}
\noindent with $\Delta T = T_{\rm h} - T_{\rm c}$, $I$ the electrical current flowing in the system, $K_0 = \kappa S / \ell$ the thermal conductance of the thermoelectric module and $R = \ell / (\sigma S)$ its electrical resistance, $S$ being the section and $\ell$ being the length of this system. The first term of the right hand side of each relation corresponds to thermoelectric convection \cite{Thomson1856} and the second term corresponds to heat conduction. While the conduction remains constant along the device, the convective contribution is modified due to the temperature variation from one edge of the system to the other: This change is the footprint of the thermoelectric conversion of energy. Besides these two terms, there is also a contribution from Joule heating associated with electrical current. The global power dissipated through Joule heating inside the system is equally released in each reservoir. Interestingly, Eqs.~(\ref{iqc}) and (\ref{iqh}) remain valid for mesoscopic structures since, in this case, equipartition of the Joule heating still holds \cite{Gurevich1997}. Note that for these mesoscopic systems, Joule heating associated with the finiteness of the quantum conductance, is not generated inside the system itself but directly into the reservoirs \cite{Datta}. However, as we focus on the heat exchanged between the system and the reservoirs, it has no consequences on the modeling of the heat currents $I_{Q,c}$ and $I_{Q,h}$.

From the electrical viewpoint, a thermoelectric module behaves as a Th\'evenin generator \cite{Apertet2012}. Indeed, the electrical current $I$ flowing from the cold to the hot side is given by:
\begin{equation}\label{eleccurrent}
I = \frac{V - \alpha \Delta T}{R},
\end{equation}
\noindent where $V$ is the voltage across the system. The thermoelectric contribution, $\alpha \Delta T$, thus appears as a counter-electromotive force in the refrigerator regime (and as an electromotive force in the generator regime) while $R$ is the internal electrical resistance.

\begin{figure}
	\centering
		\includegraphics[width=0.5\textwidth]{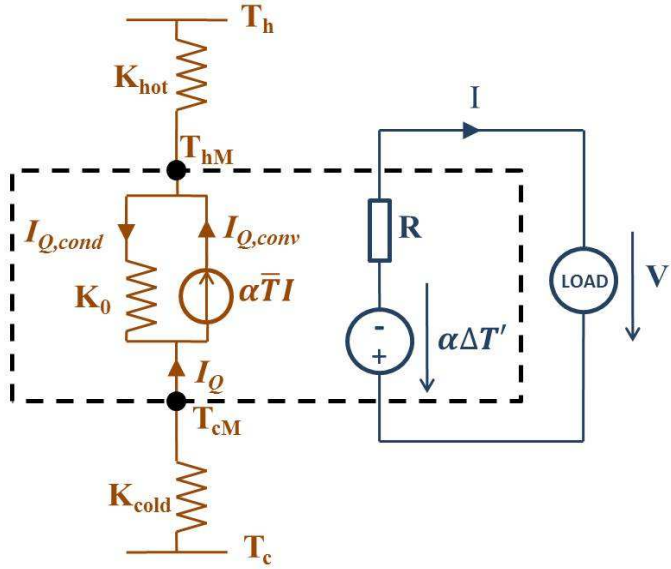}
	\caption{Schematic description of a thermoelectric refrigeration system. $I_{Q,cond}$ and $I_{Q,conv}$ are respectively the conductive and the convective parts of the heat current.}
	\label{fig:figure2}
\end{figure}

The power $P$ used to move heat from the cold reservoir to the hot reservoir is given by $I_{Q,h} - I_{Q,c}$. From Eqs.~(\ref{iqc}) and (\ref{iqh}), one gets:
\begin{equation}\label{power}
P = \alpha \Delta T I + R I^2.
\end{equation}
\noindent The first term on the right hand side of this equation reflects the power needed to work against the counter-electromotive force generated by the temperature difference while the second term is associated with power dissipation through Joule heating. Even if Joule heating is sometimes believed to pertain to nonlinear modeling of thermoelectric systems (see e.g.~\cite{Izumida2012,Izumida2015}) due to its quadratic dependence to electrical current $I$, or equivalently to the applied voltage $V$, we stress that this contribution is actually a direct consequence of the linear local description \cite{Callen1951}. As such, while it is sometimes possible to consider \emph{mathematically} linear relations neglecting Joule heating for sake of simplicity, from a \emph{physical} viewpoint, the comprehensive linear modeling of thermoelectric systems encompasses this contribution \cite{Apertet2013}. In Ref.~\cite{Fuchs2014}~, Fuchs indicates that while neglecting the contribution of Joule heating in Eqs.~(\ref{iqc}) and (\ref{iqh}) seems reasonable for thermoelectric system working as generators, it is not always the case for other regimes. Indeed, in these other cases, the electrical current might become very large. However this assumption remains acceptable for thermoelectric refrigerator as long as $I$ is much smaller than $2\alpha T_{\rm c}/R$ since in this case Joule heating is negligible compared to the convective part of the thermal current. It is then possible for convenience to consider that the heat current $I_{Q}$ remains almost constant along the system :  
\begin{equation}\label{iqsimple}
I_{Q,h} \approx I_{Q,c} \approx I_{Q} = \alpha \overline{T} I - K_0 \Delta T.
\end{equation}
\noindent This simplification amounts to neglecting both Joule heating and energy conversion but it may lead to useful results from the practical viewpoint (see e.g. \cite{Apertet2012}) even if it is not fully accurate from a thermodynamic viewpoint. Note that this approximation is also implicitly used when one considers the local definition of the heat current instead of the global one to describe the whole system (see e.g. \cite{Benenti2013}). This simplified description of the thermoelectric refrigerator is summarized on the Fig.~\ref{fig:figure2}. For the sake of completeness, we may also consider finite thermal conductances between the system and the thermal reservoirs. Their influence on the system is detailed below.

\subsection{Extended model with finite thermal contact conductances}

In order to also consider external dissipations due to non ideal contacts, we extend the previously considered model by introducing finite thermal conductances  between the system and the hot and cold reservoirs, denoted respectively $K_{\rm hot}$ and $K_{\rm cold}$, as depicted in Fig.~\ref{fig:figure2}. The main consequence of these finite thermal conductances is the modification of the temperatures at the edges of the system. The temperature difference seen by the system is thus no longer $\Delta T$. The actual temperature difference is labeled $\Delta T'$. The constitutive equations of the system, Eqs.~(\ref{eleccurrent}) and (\ref{iqsimple}), might then be rewritten as:
\begin{subequations}\label{prime}
\begin{eqnarray}
V = RI + \alpha \Delta T', \label{eq:V}\\
I_{Q} = \alpha \overline{T} I - K_0 \Delta T'.\label{iqextended}
\end{eqnarray}
\end{subequations}
\noindent Hence, it is possible to derive an analytical expression for the temperature difference $\Delta T'$: Still assuming that the heat flux remains constant along the device, a thermal equivalent of a voltage divider leads to 
\begin{equation}\label{deltaTprime}
\Delta T' = T_{hM} - T_{cM} = \Delta T + \frac{I_Q}{K_c},
\end{equation}
\noindent where $I_Q$ is given by Eq.~(\ref{iqextended}) and $K_c$ is the equivalent thermal conductance of the contacts \cite{Apertet2012,Freunek2009}:
\begin{equation}
K_c = \frac{K_{hot}.K_{cold}}{K_{hot}+K_{cold}}. \nonumber
\end{equation}
\noindent Note that in the refrigerator regime the temperature difference at the edges of the thermoelectric system $\Delta T'$ is higher than $\Delta T$.

\subsection{Block diagram description}

The block diagram representation is mainly used in automatic control engineering. As stressed by Raven \cite{Raven1961}, \emph{``these diagrams have the advantage of indicating more realistically the actual processes which are taking place, as opposed to a purely abstract mathematical representation''}. It thus appears as a way to focus on the physics of a system rather than solely on its mathematical description. Furthermore, basic rules are relatively simple: A circle is the symbol which is used to indicate summing operation while a box is the symbol for multiplication.
Each block is connected to the next block by an unidirectional arrow (this one-way relationship is particularly interesting when it comes to causality issue). More details on the block diagram representation might be found in the Raven's book \cite{Raven1961}.

\begin{figure}
	\centering
		\includegraphics[width=0.50\textwidth]{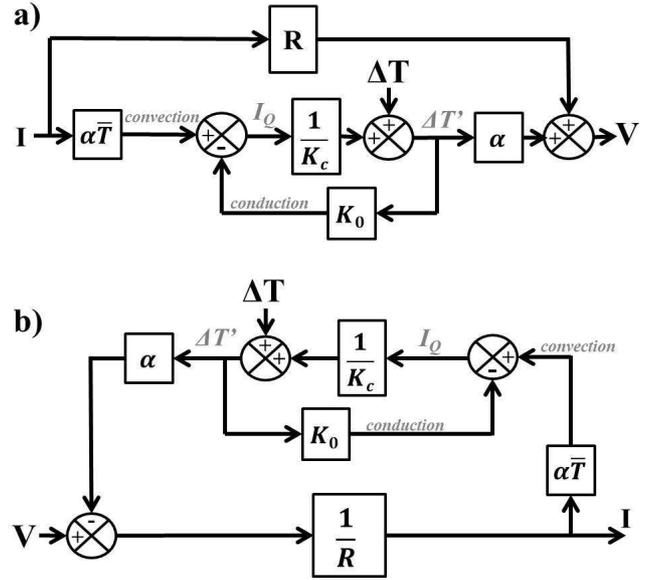}
	\caption{Block diagrams associated with a simplified thermoelectric conversion system connected to thermal reservoirs through finite thermal conductances. The term \emph{convection} corresponds to $\alpha \overline{T} I$ and the term \emph{conduction} corresponds to $K_0 \Delta T'$.}
	\label{fig:figure4}
\end{figure}
Combining Eq.~(\ref{prime}) and Eq.~(\ref{deltaTprime}), one may thus describe the thermoelectric system as a block diagram relating the electrical current $I$ to the voltage $V$ across the system (Fig.~\ref{fig:figure4}.a) or as a block diagram relating $V$ to $I$ (Fig.~\ref{fig:figure4}.b).
In the first of these diagrams, one might clearly see two distinct paths relating the electrical current $I$ to the voltage $V$: The upper path simply is the contribution of Ohm's law (first term of Eq.~(\ref{eq:V})) while the lower path reflects the dependence of the counter-electromotive force $\alpha\Delta T'$ (second term of Eq.~(\ref{eq:V})) with the electrical current $I$ through the modulation of $\Delta T'$ by the convective heat flux $\alpha \overline{T} I$. In the second diagram (Fig.~\ref{fig:figure4}.b), these two distinct paths are recovered, except that the factor associated with Ohm's law is now $1/R$ as $I$ and $V$ has been inverted. The appropriate diagram will be chosen depending on the imposed electrical constrain. This modeling is one of the main results of this article, along with its interpretation.

The use of block diagrams is further justified by the presence of feedback loops as recently discussed in Ref.~\cite{Goupil2016}: In both diagrams of Fig.~\ref{fig:figure4}, the temperature difference at the edges of the thermoelectric system $\Delta T'$ intervenes in its own computation through the conductive heat current $K_0 \Delta T'$. Similarly, Fig.~\ref{fig:figure4}.b shows that the appearance of an electrical current $I$ moderates the electromotive force $\alpha\Delta T'$ through thermoelectric convection, thus moderating the value of this same electrical current. These feedback loops are at the heart of the automatic control theory since they are used to efficiently regulate systems. Block diagram description is thus especially appropriate to highlight such mechanisms. Note however that the approach used in the present article differs slightly from the one used in Ref.~\cite{Goupil2016}: In each box of Fig.~\ref{fig:figure4}, the multiplying coefficient is constant and thus independent of the working conditions. This is not the case in Ref.~\cite{Goupil2016} as each factor depends on the electrical load resistance; Interpretation of the block diagrams thus seems more straightforward with our approach since there is no \emph{implicit} relations between variables. Furthermore, while we focus on the relation between the electrical current $I$ and the voltage $V$, Goupil and coworkers focus on the relation between the temperature difference $\Delta T$ and one of the electrical variables ($I$ or $V$).

\section{\label{dissipation} On the benefits of dissipations}

With this diagrammatic description of the thermoelectric heat engine, it is now possible to stress the need for dissipation in such a system. We distinguish in this section the usefulness (or not) of each dissipation source. 

\subsection{Driving the system}

As discussed in the Introduction, causality might be seen as the possibility of driving the engine thanks to the load. The electrical load is able to drive the system if the choice of one of the electrical parameters, i.e, the voltage $V$ or the electrical current $I$, leads to an unambiguous working point where these two parameters are clearly defined. This condition is met only if there is at least one path linking the imposed parameter to the other in the block diagrams displayed in Fig.~\ref{fig:figure4}. We show in the following that each path is actually related to a dissipation source.

\subsection{Insights from the block diagrams}

In the building blocks displayed in Fig.~\ref{fig:figure4}, each dissipation source is represented by a single box: The thermal conductance $K_0$ stands for the heat leaks through conductive process, the thermal conductance $K_c$ stands for the dissipations in the non-ideal thermal contacts and $R$ stands for Joule heating. Some blocks also involve the Seebeck coefficient $\alpha$; They are however not associated with dissipations since this coefficient, reflecting the coupling between heat flux and electrical current, on the contrary represents the useful part of the energy conversion process. For the thermoelectric Carnot engine, both $K_0$ and $R$ should vanish while $K_c$ should become infinite. In this case, Fig.~\ref{fig:figure4}.a clearly demonstrates that the imposed electrical current $I$ and the resulting voltage across the system $V$ are no more linked: As already pointed out, this engine is non causal. This analysis still holds if one turns to Fig.~\ref{fig:figure4}.b where the voltage is imposed rather than the electrical current. It is also the case in the generator regime when the load is a resistance: None of these quantities is then fixed as the working point is determined by the intersection of the characteristics of the thermoelectric system and those of the resistive load.

From this ideal situation, we can sequentially ``\emph{switch on}'' each individual dissipation source in order to assess their impact on the causality of the system:\\
- When friction is the only dissipative process, i.e. $R \neq 0$, the connection between $I$ and $V$ is straightforward. This situation is known as exoreversibility \cite{Chen1997}.\\
- When the only dissipative process results from the finiteness of the thermal conductance of the thermal contact, there still is a direct path in the block diagram between $I$ and $V$. The associated engine is thus causal. From Fig.~\ref{fig:figure4}.a, it is possible to associate this path with a virtual electrical resistance given by:
\begin{equation}\label{Rprime}
R' = \frac{\alpha^2 \overline{T}}{K_c}.
\end{equation}
\noindent This resistance has already been derived using another approach in Ref.~\cite{Apertet2012}. In Fig.~\ref{fig:figure4}.b, the condition $R=0$ implies that the electromotive force $\alpha \Delta T'$ should always be equal to the imposed voltage $V$ since the block $1/R$ demonstrates an infinite gain in this case. This behavior is similar to the one of a perfect operational amplifier with negative feedback: This feedback ensures the dependence of $I$ on $V$ and thus the causality of the system. This situation, considered in the seminal paper by Curzon and Ahlborn \cite{Curzon1975} and later applied to a thermoelectric generator \cite{Agrawal1997}, is known as endoreversibility \cite{Rubin1979}.\\
- Finally, we consider the situation where dissipations are only due to heat leaks, i.e., $K_0 \neq 0$. In this case, there is no connection between the two electrical variables: The causality is \emph{not} ensured even if there are dissipations. To this extent, heat leaks could not be considered as useful since they deteriorate performances but do not contribute to the causality of the system: They are only detrimental. This point justifies the use of the \emph{strong coupling assumption} in finite time thermodynamics and the efforts to reduce conduction in real thermoelectric systems (see e.g. \cite{Zhao2014}).

\subsection{Coefficient of performance versus cooling power characteristics}

\begin{figure}
	\centering
		\includegraphics[width=0.5\textwidth]{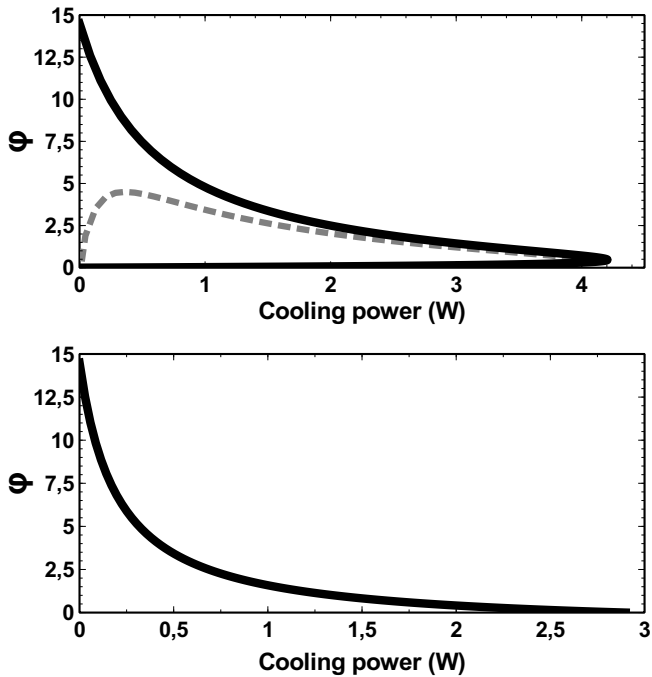}
	\caption{Coefficient of performance $\varphi$ as a function of the cooling power $I_{Q,c}$ for an exoreversible thermoelectric refrigerator (upper panel) and for an endoreversible thermoelectric refrigerator (lower panel). The dashed line in the upper panel corresponds to a system with additional parasitic heat conduction.}
	\label{fig:figure5}
\end{figure}

Following the approach of Gordon in Ref.~\cite{Gordon1991}, we also discuss the causality of a thermoelectric system focusing on the evolution of its performances when the working conditions is varied. We still consider thermoelectric refrigerators contrary to Gordon who deal with thermoelectric generators. We thus display in Fig.~\ref{fig:figure5} the characteristic curves relating the coefficient of performance $\varphi$ and the cooling power $I_{Q,c}$ rather than the power-efficiency curves used in Ref.~\cite{Gordon1991}. 
On such graphs, the plurality of the working points is the footprint of causality. These different points are reached thanks to the driving of the load. On the contrary, for a Carnot refrigerator, the working conditions are limited to a single point where the coefficient of performance is maximal, i.e., $\varphi_C = T_c / (T_h - T_c)$, but where the cooling power vanishes. Thermoelectric refrigerators with only heat leaks also display a single working point but where both the coefficient of performance and the cooling power vanish. Typical behaviors of both endoreversible and exoreversible thermoelectric systems are shown on Figure~\ref{fig:figure5}. While both refrigerators demonstrate Carnot efficiency in the limit of quasistatic state, i.e., $I=0$, there is a difference at higher electrical current: On the one hand, the exoreversible engine experiences Joule heating that compensates at some point the convective heat flux, leading the cooling power to vanish. On the other hand, the cooling power of the endoreversible engine increases continuously until the coefficient of performance vanishes. 

In a recent article \cite{Entin2014}, Entin-Wohlman and coworkers claimed that heat leaks might be crucial for the usefulness of thermoelectric devices. They argue that without such parasitic effects the cooling power at maximum efficiency vanishes, making this refrigerator quite unpractical. However, as clearly demonstrated in the present article, heat leaks could not be considered as useful since they are only detrimental from both causality and performances viewpoints. It appears that in their derivations, Entin-Wohlman and coworkers have overlooked the possibility to use the load to drive the system away from the quasistatic condition (where both cooling power and electrical current vanish). The comparison of two refrigerators with (solid line) and without (dashed line) heat leaks is displayed in the upper panel of Fig.~\ref{fig:figure5}. The endoreversible refrigerator indeed shows a vanishing cooling power at maximum efficiency contrary to the same system with heat leaks. However, there is a large range of working conditions for which the performances of the endoreversible refrigerator are much higher than that of the leaky one (see additional discussion in Ref.~\cite{Apertet2013EPL}).

\section{Conclusion}

Building on the specific case of a thermoelectric refrigerator, we have recover that the causality of an energy conversion system may result from finite thermal conductance between this system and the heat reservoirs and/or from frictions leading to Joule heating in the thermoelectric case. However, dissipations resulting from thermal conduction cannot be used to this purpose: It appears as a pure parasitic process regarding energy conversion and should thus always be avoided contrary to the claims made in Ref.~\cite{Entin2014}.

\section*{Acknowledgments}
I am pleased to thank Dr.~H.~Ouerdane for his careful reading of the manuscript and Prof.~C.~Goupil for suggesting the use of block diagrams.

\appendix
\section*{Appendix: Reaching the different regimes}

\begin{figure}
	\centering
		\includegraphics[width=0.5\textwidth]{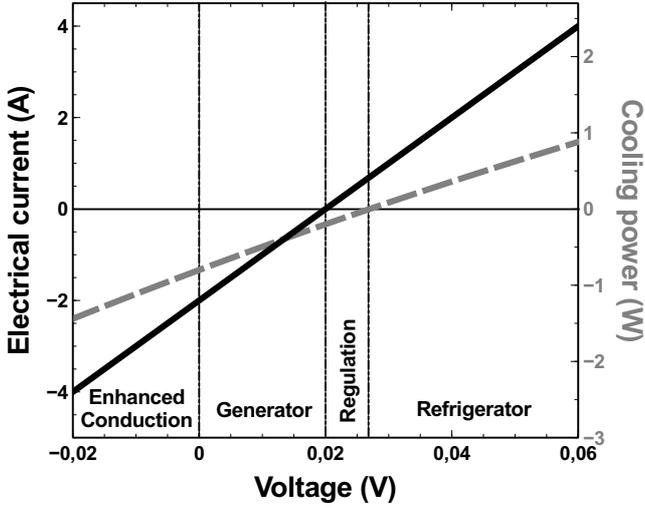}
	\caption{Electrical current $I$ and cooling power $I_{Q,c}$ as a function of the applied voltage $V$. This example corresponds to a system with the following properties: $T_{\rm h} = 310~K$, $T_{\rm c}= 290~K$, $\alpha = 0.001~V.K^{-1}$, $K_0 = 0.01~W.K^{-1}$, $R = 0.01~\Omega$ and $ZT = 3$.}
	\label{fig:figure3}
\end{figure}

While we choose to describe the system in the refrigerator regime, it is interesting to also consider other working regimes. The choice of the working condition is obtained thank to the load. It possible to distinguish four different regimes of interest depending on the voltage $V$ across the system as depicted on Fig.~\ref{fig:figure3}. When this voltage is comprised between 0 and $\alpha \Delta T$, the system works as a generator and is thus able to deliver power to the load: Since the electrical current $I$ is negative within this range, the power used $IV$ is also negative. The limiting values of $V$ correspond respectively to the short circuit and open circuit conditions. The heat current is also a function of the working conditions because of its convective part as already stressed in Refs.~\cite{Min2008,Apertet2012}. This dependence may be given through the introduction of an extended Fourier law \cite{Apertet2012}:
\begin{equation}\label{Fourier}
I_Q = - K_{\rm eff} \Delta T
\end{equation}
\noindent where the effective thermal conductance of the system $K_{\rm eff}$ is defined as:
\begin{equation}\label{Keff}
K_{\rm eff} = K_0 \left(1 - \frac{I}{I_{\rm sc}}ZT\right),
\end{equation}
\noindent with $I_{\rm sc} = \alpha \Delta T / R$ the value of the electrical current $I$ when the system is short circuited. The relation~(\ref{Fourier}) is only an other way of expressing Eq.~(\ref{iqsimple}). Thus, thermoelectric systems may be viewed as variable thermal resistor \cite{Min2008}. If the voltage applied to the system is below $\alpha \Delta T$, $K_{\rm eff}$ is higher than $K_0$, the thermal conductance associated with thermal conduction only. Indeed, in this case, the electrical current is negative and, hence, the convective part of the heat current transports heat from hot to cold just as thermal conduction does. While this additional contribution occurs naturally in the generator regime, it is necessary to provide power to the system to further increase the heat current when the voltage becomes negative. This range of working conditions is associated with an \emph{enhanced conduction}. This regime may have interesting applications in electronic cooling as recently discussed by Zebarjadi \cite{Zebarjadi2015}.  

When the voltage is increased above its open-circuit value, i.e., $\alpha \Delta T$, the electrical current becomes positive. Even if the conductive part of the heat current then transports heat from cold to hot at the expense of electrical power consumption, the refrigerator regime is not reached immediately. Indeed, the losses from hot reservoir to cold reservoir through heat conduction, $K_0 \Delta T$, have to be fully compensated first. Consequently, there is an intermediate regime where the cooling power $I_{Q,c}$ remains negative even if electrical power is consumed. In this range, one may modulate the effective thermal conductance $K_{\rm eff}$ through the electrical working conditions as discussed recently by Colomer et al. \cite{Colomer2015}. It is even possible, in theory, to reach thermal insulation with this method. This regime has been labeled as \emph{regulation} since it is then possible to regulate heat current through thermoelectric convection \cite{These}.

The cooling power is obviously limited in real systems: At some point, the contribution of the Joule heating is no longer negligible and one has then to consider Eq.~(\ref{iqc}) instead of Eq.~(\ref{iqsimple}). When the voltage is high enough, Joule heating becomes the paramount contribution to the heat current and this latter is then negative. Note that the working condition associated with a vanishing heat current, i.e., when the convective current is exactly compensated by both the conductive current and the Joule heating, is the one used to determine the maximum temperature difference reached by a thermoelectric refrigerator \cite{Ioffe1957}.

\end{document}